\documentclass[
atoms,article,
accept,moreauthors,
10pt,a4paper]{mdpi} 

\usepackage{hyperref} 
\usepackage{subcaption}

\firstpage{1} 
\articlenumber{x}
\doinum{10.3390/------}
\pubvolume{xx}
\pubyear{2018}
\copyrightyear{2018}
\externaleditor{Academic Editors: Ivette Fuentes, Philippe Bouyer}
\history{Received: date; Accepted: date; Published: date}

\global\long\def\gbar{\ensuremath{\bar{g}}}

\pdfoutput=1

\Title{\rightline{{\rm\small IIT-CAPP-17-5}} Studying Antimatter Gravity with Muonium}

 \Author{MAGE Collaboration: Aldo~Antognini~$^{1,2,\dagger}$, Daniel~M.~Kaplan~$^{3,*,\dagger}$, Klaus~Kirch~$^{1,2,\dagger}$, Andreas~Knecht~$^{1,\dagger}$, Derrick~C.~Mancini~$^{3,\dagger}$, James~D.~Phillips~$^{3,\dagger}$, Thomas~J.~Phillips~$^{3,\dagger}$, Robert~D.~Reasenberg~$^{4,\dagger}$, Thomas~J.~Roberts $^{3,\dagger}$ and Anna Soter $^{1,\dagger}$ 
 }

\AuthorNames{Daniel M Kaplan, Klaus Kirch, Derrick C Mancini, James D Phillips, Thomas J Phillips, Robert D Reasenberg, Thomas J Roberts, and Anna Soter}

\address{%
$^{1}$ \quad Paul Scherrer Institute, 5232 Villigen, Switzerland\\
$^{2}$ \quad ETH Z\"{u}rich, 8092 Z\"{u}rich, Switzerland\\
$^{3}$ \quad Illinois Institute of Technology, Chicago, IL 60616, USA\\
$^{4}$ \quad Center for Astrophysics and Space Sciences, University of California at San Diego, La Jolla, CA 92093, USA }

\corres{Correspondence: kaplan@iit.edu; Tel.: +1-312-567-3389}

\firstnote{These authors contributed equally to this work.}

\abstract{The gravitational acceleration of antimatter, \gbar, has yet to be directly measured; an unexpected outcome of its measurement could change our understanding of gravity, the universe, and the possibility of a fifth force.  
Three avenues are apparent for such a measurement: antihydrogen, positronium, and muonium, the last requiring a precision atom interferometer and novel 
 muonium beam under development. The 
interferometer and its few-picometer  alignment and calibration systems appear feasible. With 100\,nm grating pitch, measurements of \gbar\ to  10\%, 1\%, or better  can be envisioned. These could constitute the first gravitational measurements of leptonic matter, of 2$^{\rm nd}$-generation matter, and possibly,
of antimatter.
}


\begin{document}


\section{Introduction}

The question of antimatter gravity, first raised in the 1950s~\cite{Morrison}, is  of continuing interest~\cite{Nieto-Goldman,Fischler-Lykken-Roberts}. To date, decades of experimental effort 
have yet to yield a statistically significant direct measurement. Antimatter gravity studies using antihydrogen are ongoing~\cite{ALPHA,AEGIS,gbar}, and experiments with positronium have been discussed~\cite{Cassidy-Hogan}. We report here on progress towards  a measurement with muonium (Mu), an exotic atom consisting of an electron bound to an antimuon.

The most sensitive limits on antimatter gravity currently come from {\em indirect} tests  (for example, equivalence principle tests using torsion pendula~\cite{EotWash} or masses in Earth orbit~\cite{MICROSCOPE}), relying on the expected amounts of virtual antimatter in the nuclei of various elements. Stringent, albeit model-dependent, limits have thereby been set on the gravitational acceleration, \gbar, of antimatter on Earth (e.g., $\sim$\,10$^{-7}$~\cite{Alves}). The virtual muon--antimuon component of nuclear wave functions being  negligible, the extent to which these indirect limits apply to muonium is  far from obvious. Another limit,  $|\alpha_g-1|<8.7\times10^{-7}$~\cite{Ulmer}, has been derived from the measured cyclotron frequency of magnetically confined antiprotons, compared with that of H$^-$ ions, based on the gravitational redshift due to Earth's gravitational potential in the field of the local galactic supercluster~\cite{Hughes,Good}; it too need not apply to antimuons.\footnote{And we note that arguments based on absolute gravitational potentials have been critiqued by Nieto and Goldman~\cite{Nieto-Goldman}.} A {\em direct} test of the gravitational interaction of antimatter with matter is desirable on quite general grounds~\cite{Nieto-Goldman}.\footnote{The only published direct test so far~\cite{ALPHA} has yielded the limit  $-65 < \gbar/g < 110$.} Such a measurement can be viewed as a test of general relativity 
or as a search for a fifth force and is of interest from both perspectives. 

Although the equivalence principle experiments 
indicate that nuclear binding energy gravitates in the same way as ordinary mass, absent validated models of gravity at a subnuclear scale, it is unclear how the gravitational interactions of virtual  
matter should be treated. Use of a pure-leptonic atom, such as positronium or muonium, evades these complexities.  Moreover, no measurement has yet been made of the gravitational force on second- or third-generation matter or antimatter (although, with some assumptions, stringent limits can be obtained from neutral-meson oscillations, especially for $K^0$--${\overline K}{}^0$~\cite{Karshenboim}).
Since direct gravitational measurements on other higher-generation particles, such as hyperons, $\tau$ leptons, and $c$ or $b$ hadrons, appear impractical due to their short lifetimes, muonium may  be the only access we have. Recent work~\cite{Kostelec-Tasson, Tasson,Tasson2}
examining a possible standard-model extension 
 emphasizes the importance of
second-generation gravitational measurements. 
Current interest in ``fifth force'' models~\cite{Glashow-etal,Hill} (stimulated by evident anomalies in the leptonic decays of $B$ mesons) also supports more detailed investigations of muonium.

General relativity (GR) is generally taken to predict identical behaviors of antimatter
and matter in a gravitational field. 
 With the  observation of gravitational waves~\cite{LIGO}, most of the predictions of GR  are now experimentally  confirmed. Nevertheless, GR is fundamentally
incompatible with quantum mechanics, and the search for a quantum theory of gravity continues~\cite{QG}. To date, the experimental evidence on which to base such a theory comes from observations of matter--matter and matter--light interactions. In a quantum field theory, matter--matter and matter--antimatter forces can differ\,---\,for example, suppressed scalar and
vector terms might cancel in matter--matter interactions, but add in
matter--antimatter ones~\cite{Nieto-Goldman}, leading to small equivalence principle violations.
Matter--antimatter
measurements could thus play a key role. 

While most physicists expect that the equivalence principle applies
equally to antimatter and to matter, this has yet to be experimentally verified. Moreover, theories in which this symmetry
is maximally violated, effectively giving antimatter negative gravitational mass,\footnote{For example, some authors~\cite{Kerr-Newman1,Kerr-Newman2,Kerr-Newman3,Kerr-Newman4} have suggested that antimatter be identified with the $r<0$ solutions of GR's Kerr--Newman equation; in this case $\gbar=-g$ would be the expected GR result.} are
attracting increasing interest~\cite{Kowitt,Blanchet-LeTiec-2008,Burinskii,Blanchet-LeTiec-2009,Hajdukovic-2011,Villata-2011,BenoitLevy-Chardin,Hajdukovic-2012,Villata-2012,Villata-2013}
as potentially providing alternatives to cosmic inflation, CP violation, dark matter, and dark energy in explaining six great mysteries
of physics and cosmology~\cite{QuantumUniverse}: (1)~Why is the cosmic microwave background radiation
so isothermal~\cite{CMBtemp}? (2)~Why is the universe so flat~\cite{,QuantumUniverse,Peacock}? (3)~What happened to the antimatter produced in the Big Bang~\cite{QuantumUniverse}?
(4)~What holds galaxies together~\cite{QuantumUniverse}? (5)~Why are distant Type Ia supernovae so dim~\cite{QuantumUniverse,Riess-etal1998,Perlmutter-etal1999}?
 and (6), Why are the oldest stars older than the  predicted age of a $\Lambda=0$ universe~\cite{OldGlob}? In addition, gravitational masses of opposite signs for matter and antimatter would mean that virtual particle--antiparticle pairs would not contribute to gravitational mass, thus evading the indirect limits on antimatter gravity even for antihydrogen.\footnote{This also suggests a solution to what has been called ``the worst prediction in physics'': that the gravitational zero-point energy of the universe seems to exceed the size of the cosmological constant by a factor $\sim10^{120}$~\cite{Weinberg}.} 
The forgoing considerations provide more than sufficient motivation (especially given the relatively modest level of experimental effort and expense required) for a measurement of muonium gravity to be pursued.
 
\section{Method}
The measurement requires a new muonium beam that is under development, a precision atom interferometer, and a low-background muonium detection system. 
The experimental challenge stems from the short muon (and muonium) lifetime: $\tau_{Mu}\stackrel{.}{=}\tau_\mu=2.2\,\mu$s at rest~\cite{Mu-lifetime}. In one lifetime, if it obeys the equivalence principle, a muonium atom initially at rest thus falls by $\Delta y=\frac{1}{2}g\tau_{Mu}^2=24$\,pm\,---\,at the edge of measurability using modern atom-interferometry techniques. Fortunately, 
the exponential decay of the muon affords a statistics--precision tradeoff: the gravitational deflection increases as $t^2$, while the number of events decreases as $\exp(-t/\tau_\mu)$. The statistical optimum is thus to measure after two lifetimes; with systematic uncertainties  taken into account, the optimal time will be somewhat greater than this. This leads to a deflection to be measured of $\approx$\,0.2--0.5\,nm, which we argue is within the state of the art. 

\begin{figure}[t]
\rightline{\includegraphics[width=.3\textwidth,trim=5 450 -185 0 mm,clip]{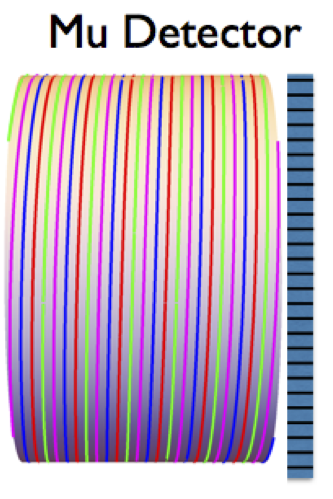}}}{\centerline{{\includegraphics[width=.6\textwidth,trim=0 0 0 -5 mm,clip]{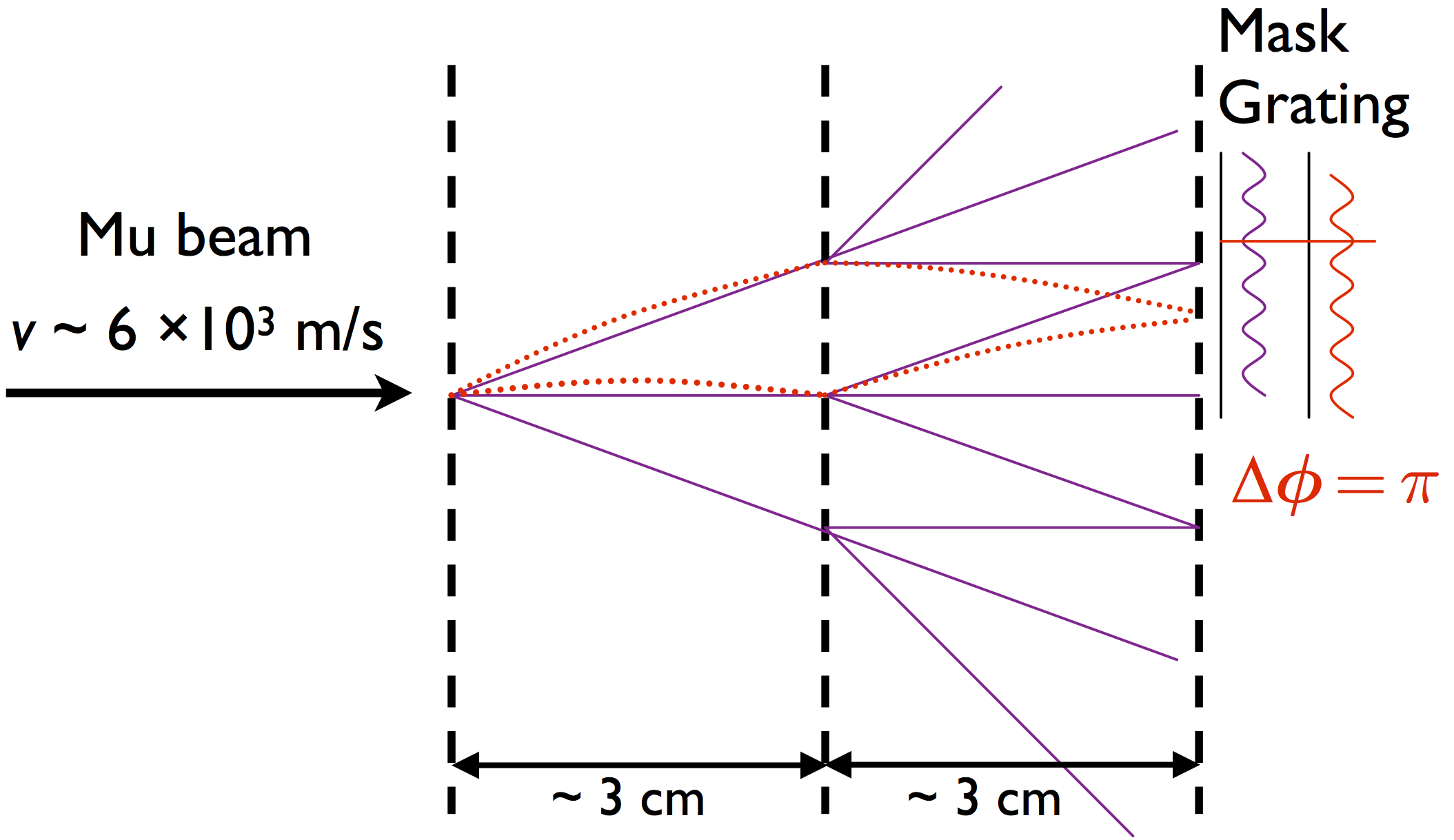}
\includegraphics[width=.28\textwidth,trim=5 35 0 55 mm,clip]{MAGE-Detector}}}
\caption{Principle of three-grating muonium-gravity interferometer, shown schematically in elevation view, with gravitational deflection and phase shift  $\Delta\phi$ exaggerated for clarity. Muonium beam enters from  left, 
slow-electron detector is at right. Not shown: ring electrodes, which accelerate slow electrons onto their detector, starting downstream of grating 3 and continuing within (helically wound) scintillating-fiber-barrel positron-tracking detector; and scintillating-bar hodoscope surrounding  positron-tracking detector.}
\label{fig:MZ}
\end{figure}

The precision we require can be achieved using a three-grating interferometer in a Mach--Zehnder-like 
arrangement (Fig.~\ref{fig:MZ}),\footnote{While we will not have the separated beams typical of Mach--Zehnder interferometers, as shown below, we expect quite visible interference fringes nonetheless~\cite{Hammond,McMorran-Cronin}. This geometry is often referred to as a Talbot interferometer by the X-ray optics community.} 
the two inter-grating separations being equal~\cite{Phillips,Kirch}.  The sensitivity of such an experiment is estimated as~\cite{Oberthaler}:
\begin{equation}\label{eq:sens}
\delta g = \frac{1}{C\sqrt{N}}\frac{d}{2\pi}\frac{1}{t^2}\,,
\end{equation}
where $C$ = 0.1 is the (conservatively estimated) fringe contrast (see below), $N$ the number of events detected, $d$ the grating pitch, and $t$ the Mu transit time between gratings.  While Eq.~\ref{eq:sens} indicates that finer grating pitch would be helpful, we have chosen $d=100$\,nm as a compromise between sensitivity and systematic error due to grating variations over the $\sim$\,cm$^2$ grating area. At a nominal rate of $10^5$ incident Mu/s, and taking into account decays and estimated inefficiencies, the statistical measurement precision is $\approx$\,0.3$g$ per $\sqrt{N_d}$, $N_d$ being the exposure time in days. 

Although conceptually similar to previous atom-beam interferometers~\cite{Keith-etal,Hammond,Brezger-etal}, the proposed precision and source size call for improvements beyond those previously built. We note that, assuming (for the sake of illustration) $\gbar = g$ and 2-lifetime grating separation, the interferometric gravitational phase difference  $\Delta\phi = \pi \gbar t^2/2d=3$\,mrad corresponds to $gt^2 \approx 0.2$\,nm of gravitational deflection. A simple way
to see this is to consider that from grating 1 to 2, the deflection is $y_2 = -\frac{1}{2}gt^2$, and at grating 3, 
it is four times as large, $y_3 = 4y_2$. The phase difference at grating 3 with respect to the diffracted beams from grating 2, which start at $y_2$, and at grating 3 have fallen an additional distance $y_2$, thus corresponds to $y_3 - 2y_2 = 190$\,pm of gravitational deflection, where $t = 2\tau_\mu = 4.4\,\mu$s.\footnote{An alternative derivation in an accelerated reference frame, in which the beam travels in a straight, horizontal line and the interferometer accelerates upwards at $g$, may be somewhat easier to follow and leads to the same conclusion.} The gratings' relative vertical alignment should therefore to be controlled with an accuracy better than 10\,pm (1\,pm) for a 10\% (1\%) measurement.

We are engaged in R\&D to develop the needed instrumentation and demonstrate the feasibility of the technique, with  three focus areas:
\begin{enumerate}[leftmargin=*,labelsep=4.9mm]
\item	Development of improved low-velocity muon and muonium beams;
\item Development of a sufficiently precise interferometer; and
\item	Development of a sufficiently precise interferometer alignment and calibration technique.
\end{enumerate}
The success of this work will culminate 
in MAGE: the Muonium Antimatter Gravity Experiment.

\subsection{Interferometer}

In a Mach--Zehnder atom interferometer, the de~Broglie wave corresponding to each incident atom diffracts at the first grating, and again at the second, interfering with itself at the third grating. This produces a sinusoidally varying image at the location of the third grating, with period equal to the grating pitch. As illustrated in Fig.~\ref{fig:MZ}, the image is slightly displaced vertically due to the gravitational deflection of the muonium atoms. With 100\,nm grating pitch, the image is too fine to be captured by a detector, but by scanning the third grating vertically using piezoelectric actuators, the image is translated into the time domain as a sinusoidally varying beam intensity. With $\approx$\,50\%-open gratings,\footnote{For overlapped interferometer beams as in our case, the optimal open fractions have been shown to be (0.60, 0.43, 0.37)~\cite{Hammond}, which can feasibly be fabricated in our proposed approach.} even orders (except zero) are suppressed, and the three diffraction orders shown contain most of the transmitted intensity. Since each atom's de Broglie wave interferes with itself, and the interference patterns from all atoms are in phase with each other, this configuration accommodates an extended, incoherent source, easing alignment and beam requirements~\cite{Cronin-etal,Chang-etal,McMorran-Cronin,Hornberger-etal}. We have modeled the performance of such an interferometer using the procedure of Refs. \cite{Ekstrom,Hammond} and find a maximum expected contrast of 20\% for the case of overlapping beams. This will of course be degraded in practice due to such factors as grating imperfections, thus the 10\% contrast assumed above is plausible.

To accommodate the anticipated beam size and divergence in various muonium source configurations, the grating area should be $\sim$\,cm$^2$ and the open fractions as close to the optimal values as practicable. The structures would ideally be grids of ~$\approx$\,1\,cm-wide horizontal slits. For mechanical stability, however, buttresses between lines of the grating are required, producing an array of mini-slits that form a grid approximating an ideal grating. The grid structures will be made in a thin film of Si$_3$N$_4$ or ultrananocrystalline diamond (UNCD), coated with higher-$Z$ metal, as membrane windows on silicon wafers. The fabrication techniques required are $e$-beam lithography and reactive-ion 
and wet etching, all available to us via our collaboration with the Center for Nanoscale Materials (CNM) at Argonne National Laboratory, as well as within the Laboratory for Micro- and Nanotechnology at the Paul Scherrer Institute (PSI). 
The CNM has developed specific process technology that has been demonstrated in UNCD and Si$_3$N$_4$ for membrane aperture structures and in UNCD for zone-plate structures very similar to that needed for the grid structures we require~\cite{CNM-struct,Zone-Plates}. 
\begin{figure}[tb]
\centering
\begin{subfigure}{0.59\textwidth}
\includegraphics[width=.99\textwidth,trim=7 10 7 0 ,clip]{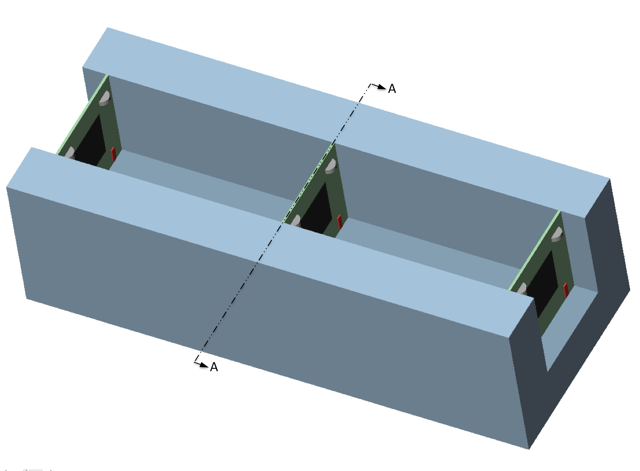}
\caption{Optical bench with gratings, showing mirrors at upper grating corners for alignment interferometers.}
\vspace{-.2in}\end{subfigure}
\begin{subfigure}{0.4\textwidth}
\includegraphics[width=.99\textwidth]{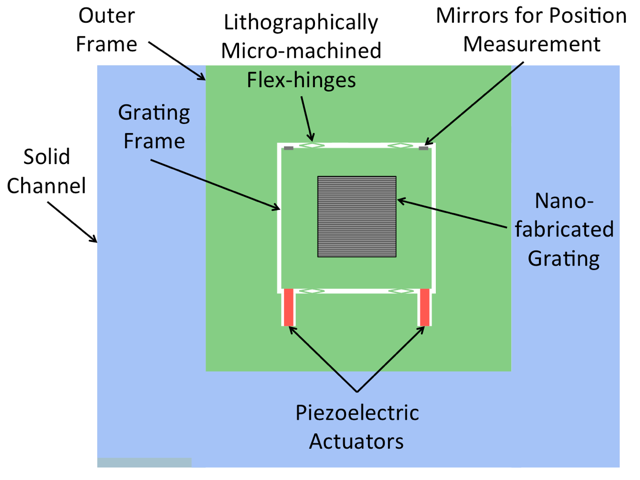}
\caption{Section A-A detail showing mounting scheme of each grid within its silicon frame.}
\end{subfigure}
\caption{Muonium interferometer support scheme in channel-cut single-crystal silicon optical bench.}\label{fig:interf}
\end{figure}

\begin{figure}
\centerline{\includegraphics[width=.7\linewidth, trim=0 0 2 0,clip]{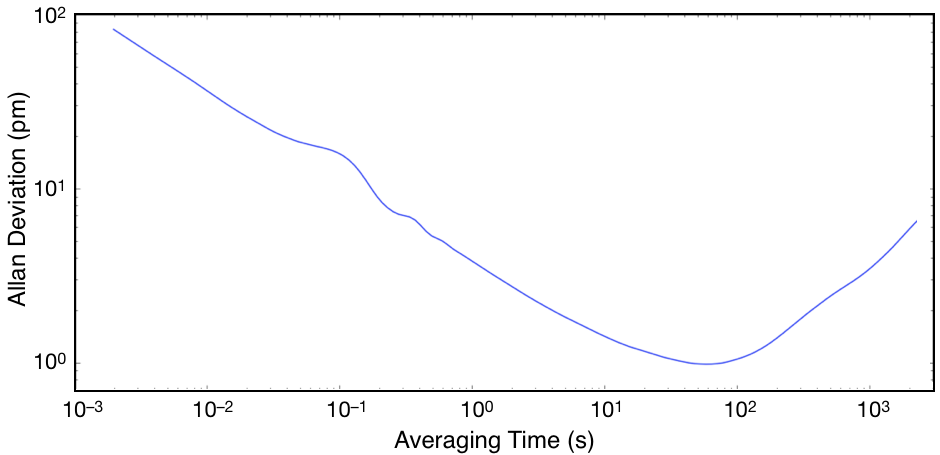}}
\caption{Semiconductor-laser tracking frequency gauge (TFG) precision in picometers vs.\ averaging time in seconds, determined by measuring the half-meter optical path-length difference of a Michelson interferometer using two TFGs  simultaneously, operated at 1,560\,nm with a common optical path (from \cite{TFG-IIT}). The demonstrated stability (Allan deviation less than 3\,pm for averaging times ranging from about 1 second to about 1,000 seconds) implies that X-ray calibrations need not be repeated more often than every 10 to 20 minutes; given operation in a cavity of modest finesse (as in Fig.~\ref{fig:optics}), the calibration interval extrapolates to 10 days.}\label{fig:Allan}
\end{figure}

\begin{figure}
\centerline{\includegraphics[width=.7\linewidth]{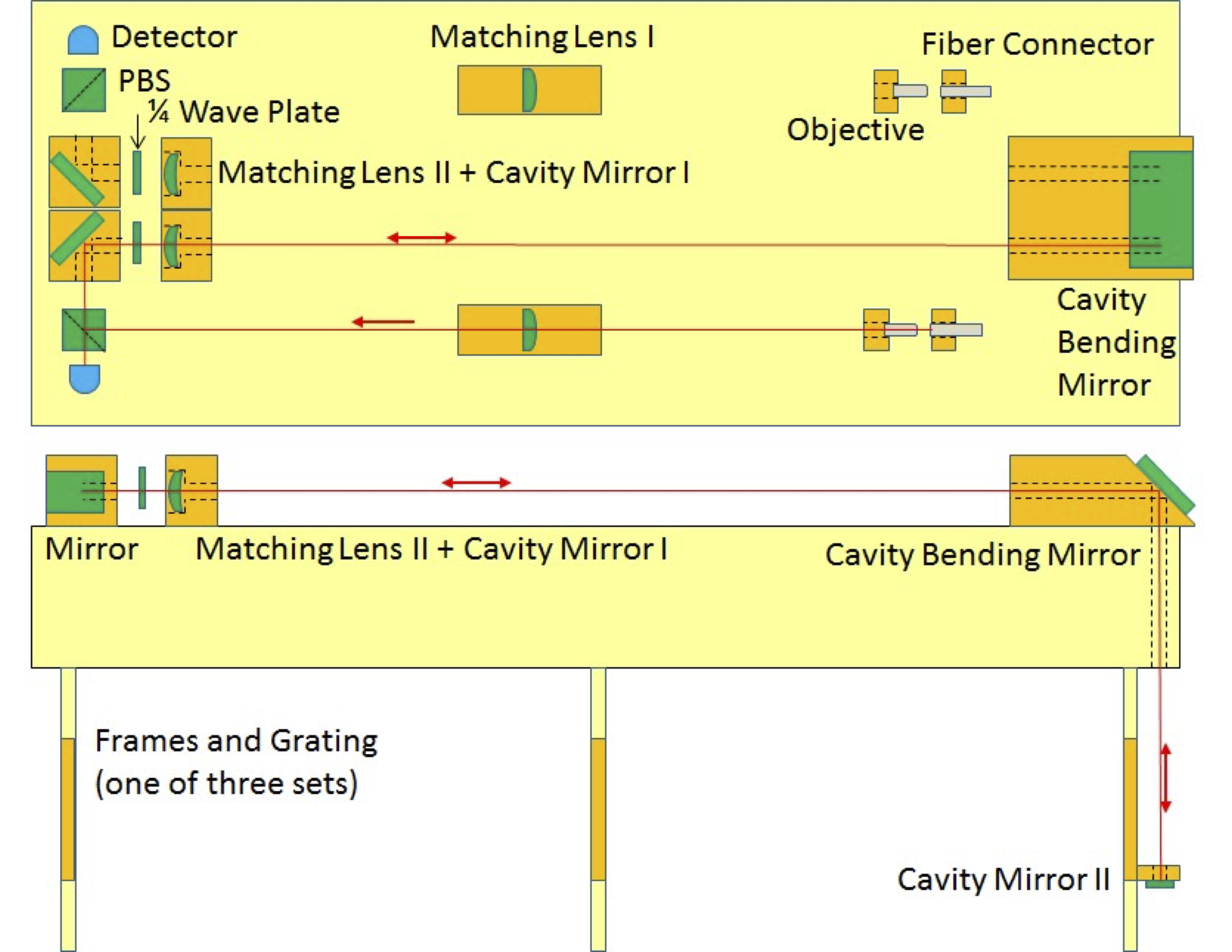}\includegraphics[width=.3\linewidth,trim=200 0 310 200,clip]{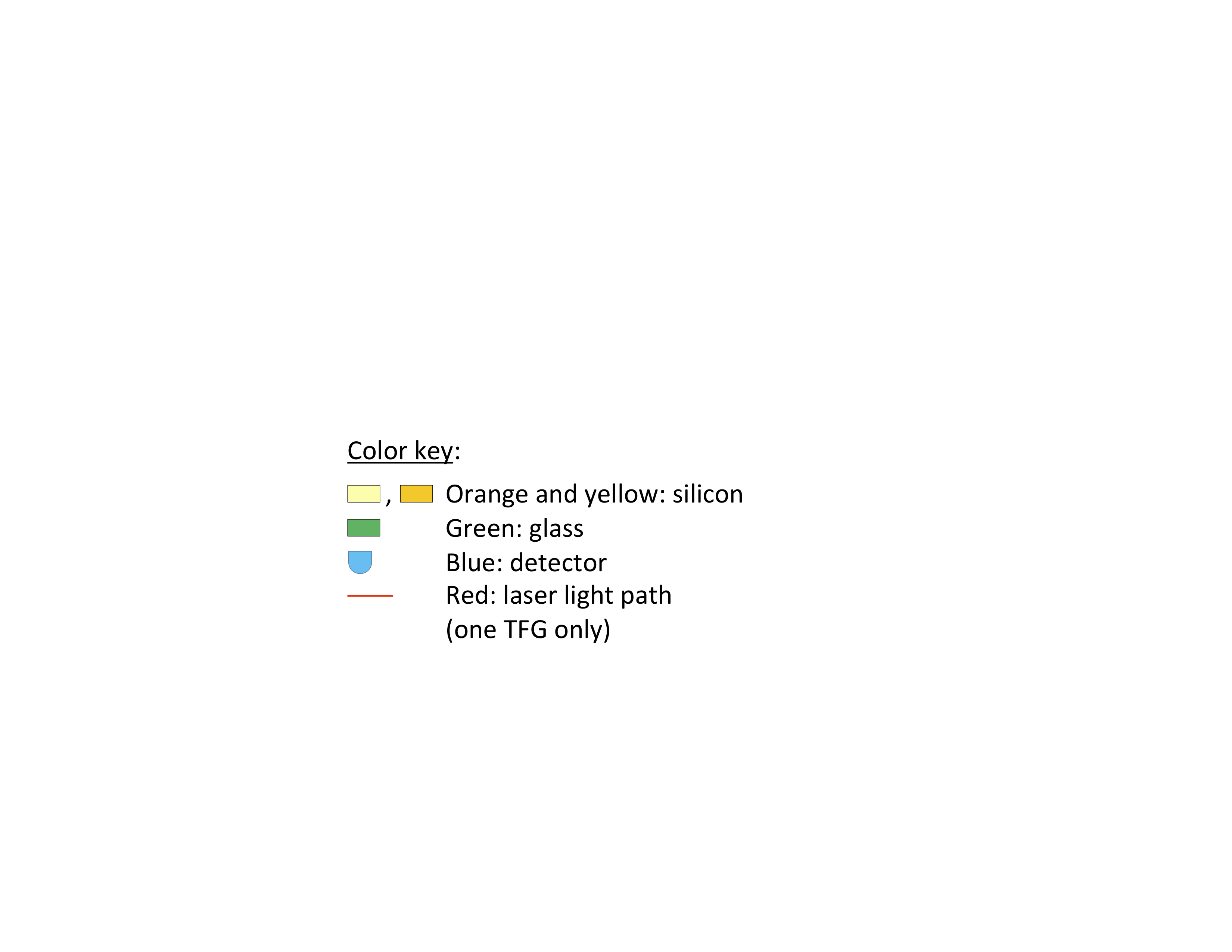}}
\caption{Optical layout concept of the two TFG measurement interferometers  observing the position of grating 3, mounted on bottom outer surface of optical bench, in two orthogonal views, with one light path indicated in red. (Similar arrangements on the outer sides of the optical bench can monitor the alignment of the other two gratings.) The TFG is shown operating in a cavity, which enhances the performance over operation in a Michelson interferometer.}\label{fig:optics}
\end{figure}

Given the small deflections to be measured, mechanical and thermal stability of the interferometer are paramount. To that end, the gratings will be mounted on an extremely stiff, channel-cut single-crystal silicon optical bench (Fig.~\ref{fig:interf}) like those developed for use in cryogenic monochromators at synchrotron radiation facilities~\cite{monochromator}, and operated at a temperature at which the thermal expansion coefficient of silicon is near zero~\cite{Swenson}:  $T\stackrel{<}{_\sim}$\,4\,K (or, for initial testing, the more convenient  $T\approx$\,125\,K). Sub-10\,pm grating alignment is also required, which is challenging using established techniques. We have therefore proposed to use the semiconductor-laser tracking frequency gauges (TFGs) recently developed for other equivalence-principle tests~\cite{Thapa-etal,Reasenberg-etal,Reasenberg-etal2}. These employ Pound--Drever--Hall locking~\cite{PDH} of a laser to the length of an interferometer (which can be nonresonant or resonant, i.e., a cavity), such that small changes in length are translated into shifts of a laser frequency. The beat frequency between the measurement laser and a reference laser locked to a stable length can be measured to high precision.
 A precision of 2\,pm has  been demonstrated using a non-resonant Michelson interferometer, and 46\,fm using a Fabry--Perot cavity with a finesse of 130~\cite{Thapa-etal}. More recently~\cite{TFG-IIT}, a precision of 1\,pm was demonstrated by some of us (D.M.K., J.D.P., T.J.R., and R.D.R.) at Illinois Institute of Technology (IIT) using the same non-resonant Michelson interferometer (Fig.~\ref{fig:Allan}). Figure~\ref{fig:optics} shows a possible arrangement of the optics that form the laser interferometers used to monitor the grating alignment.

\begin{figure}
\centerline{\includegraphics{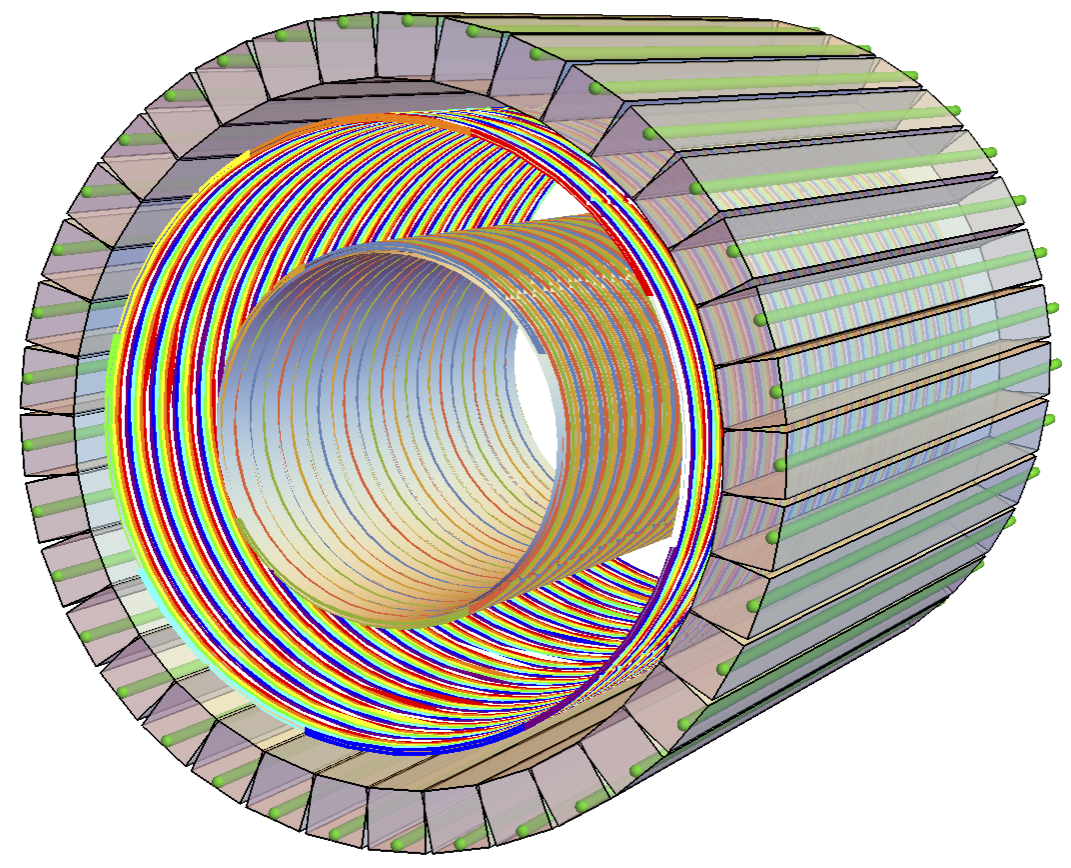}}
\caption{3D drawing of 2-layer barrel scintillating-fiber tracker, surrounded by outer scintillator-bar hodoscope used for trigger purposes and to break reconstruction ambiguity.}\label{fig:barrel}
\end{figure}

To suppress background in the muonium detector, a coincidence technique will be used. Decaying muons   emit a fast positron, leaving behind the (now unbound) electron, which can be accelerated electrostatically onto a microchannel plate (MCP) or pixel detector and counted~\cite{Matthias-etal}. Due to the $Q$ value of the $\overline{\mu}\to \overline{e}\nu \overline{\nu}$ decay, the fast positrons typically emerge at a substantial angle to the beam direction, so can be efficiently detected in a barrel scintillating-fiber tracking detector downstream of the third grating (Fig.~\ref{fig:barrel}). The detection in coincidence of an accelerated electron and a fast positron with track pointing back to the ``decay region'' unambiguously signals a muonium decay in that region, which extends  from the third grating to the slow-electron detector. To minimize the electronics channel count, the scintillating fibers are laid out as helices and read out on both ends, so that the end-to-end amplitude and time differences  can be used to localize the hit to one turn of the helix. The surrounding scintillating-bar hodoscope (with wavelength-shifting fibers for readout) is used  to determine the azimuthal angle $\phi$ of the hit, allowing determination of $z$ to about a fiber width (1\,mm). 
G4beamline~\cite{G4BL} simulations of the muonium detector indicate a $\approx$\,70\% acceptance for the coincidence requirement, which can be increased if desired by extending the barrel detector further upstream or downstream. (Due to the electrostatic acceleration of the electron, its detection efficiency is near 100\%.)

\subsection{Muonium Beam}

\begin{figure}
\centering
\includegraphics[width=.6\textwidth]{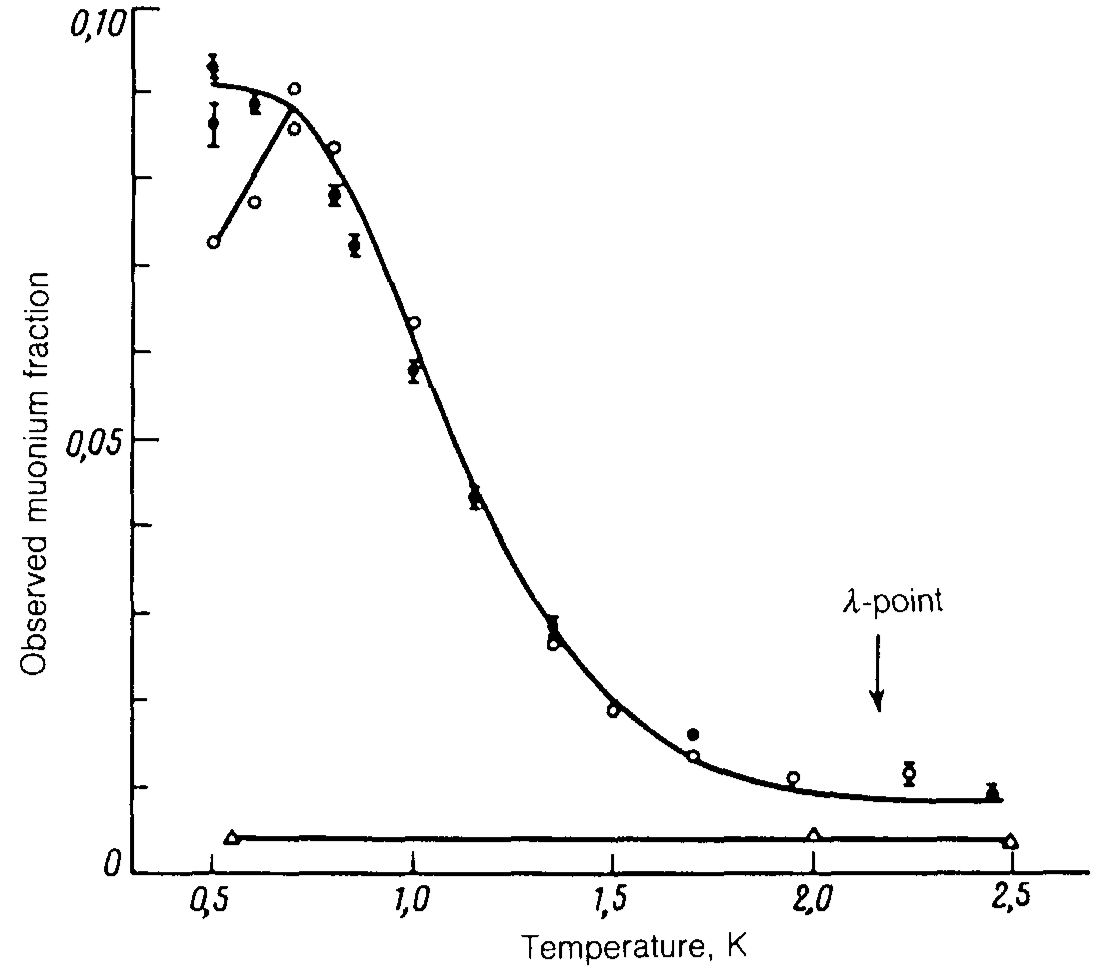}
\caption{Muonium production vs.\ temperature in liquid helium (from~\cite{Abela-etal});  open circles are results in pure $^4$He, filled circles, $^4$He + 0.2\% $^3$He, and  triangles, pure $^3$He. (Quantity plotted is observed muon decay asymmetry; an observed asymmetry of 0.105 corresponds to 100\% muonium formation.)}\label{fig:SFHe}
\end{figure}

We have identified two approaches for producing the needed muonium beam, 
both of which employ a superfluid He (SFHe) $\mu^+\to$\,Mu converter. 
SFHe has been shown to be a highly efficient converter of stopping positive muons to muonium (Fig.~\ref{fig:SFHe})~\cite{Abela-etal}; indeed, based on their experimental results, the Ref.~\cite{Abela-etal} authors conclude that below 0.7\,K essentially all $\mu^+$ stopping in helium form muonium. 
The interferometry requirements dictate the properties needed for the beam: it should be traveling in vacuum in a region free of electric and magnetic fields. Since it is neutral, and has an electric polarizability similar to that of hydrogen (among the smallest   of any atom), Mu is intrinsically insensitive to such fields; nevertheless, for the effects of gravity to stand out, even small, higher-order electromagnetic effects should be suppressed.  To optimize the stability of the interferometer structure, the Mu atoms should be moving slowly, such that the distance traveled in a few Mu lifetimes is $\stackrel{<}{_\sim}$\,10\,cm, and with a small enough range in velocity that the interference phase is not significantly smeared out. 
Because the loss of contrast  is given by $C/C_0=-\exp{(\sigma_\phi^2/2)}$~\cite{Cronin-etal}, and 
the RMS width $\sigma_\phi$ of the gravitational phase is quite small, in practice we do not expect the Mu velocity range to contribute significantly to the uncertainty of the measurement.

The first muonium-beam approach is to use the ``muCool''  $\mu^+$ beam under development by some of us (A.A., A.K., K.K., T.J.P., and A.S.)~\cite{muCool,muCool2,muCool3,Eggenberger-thesis}. In this scheme (due to D. Taqqu~\cite{Taqqu-PRL}), the surface-muon beam is cooled by a factor $\sim10^{10}$ in phase-space density prior to stopping, while losing only 3 orders of magnitude in intensity, thus resulting in an improvement of $10^7$ in brightness. This reduces the spot size to $<$\,1\,mm and dramatically enhances the stopping fraction. The muCool beam is stopped in a micron-thin film of superfluid helium (SFHe) coating the bottom of the cryostat. Mu is formed in the SFHe, and half of it diffuses to the upper surface, where (since it behaves chemically like a light hydrogen isotope), due to the negative chemical affinity of hydrogen in SFHe~\cite{Reynolds-etal,
Krotscheck-Zillich}, it is expelled into vacuum. The resulting Mu beam will be very nearly parallel (divergence $\approx$\,27\,mrad)  and monoenergetic ($\Delta E/E\approx0.1$\%), with a velocity (6.3\,mm/$\mu$s) determined by the chemical potential of Mu in SFHe~\cite{Taqqu-PRL}, whose value (270\,K)~\cite{Taqqu-PhysProc} is the largest for any hydrogen isotope. (Because the Mu atom is in thermal equilibrium with the superfluid helium prior to being ejected, both the energy spread and the angular divergence of the Mu beam are determined by the ratio of the SFHe temperature, here taken to be 0.2\,K, to the chemical potential.) The resulting interferometer acceptance is of order unity,  leading to \gbar\ sign determination with about one month's worth of beam. 

The second option exploits another idea of Taqqu's~\cite{Taqqu-PhysProc}: use a thicker ($\sim$\,100\,$\mu$m) SFHe converter with an uncooled subsurface-muon beam, giving a \gbar\ sign determination with two days' worth of beam. Since it avoids the beam loss inherent in the muCool approach, the thick-film approach could enable a  $\stackrel{<}{_\sim}$\,10\% measurement of \gbar\ in a month of beam time, again assuming adequate control of systematic uncertainties, and possibly a $\stackrel{<}{_\sim}$\,1\% measurement in a future, upgraded surface-muon beam. Since only Mu atoms formed close to the upper surface will emerge upwards from the helium to form the desired beam, an electric field is maintained in the helium (via a pool of negative charge on the upper surface of the SFHe layer) to cause the stopping $\mu^+$ to separate from their ionization trails and  drift to the upper surface, where they then form muonium. Some deadtime may be incurred if the pool of charge needs to be periodically replenished (e.g., via a submerged emitter tip). The $\sim$\,cm-wide beam results in some acceptance loss if cm-wide gratings are employed, thus larger gratings (if feasible) could be beneficial; alternatively, the SFHe deflector (described next) could have a curved surface so as to produce some focusing of the beam into the interferometer~\cite{SFHe-Lens}. In addition to providing higher beam intensity, the thick-film approach eliminates the technical risk associated with muCool, although it has other uncertainties, such as those associated with creating and maintaining the charge pool. Clarifying these issues will require further muonium-beam R\&D at PSI.

\begin{figure}[tb]
\centering
\begin{subfigure}{0.33\textwidth}
\includegraphics[width=\textwidth]{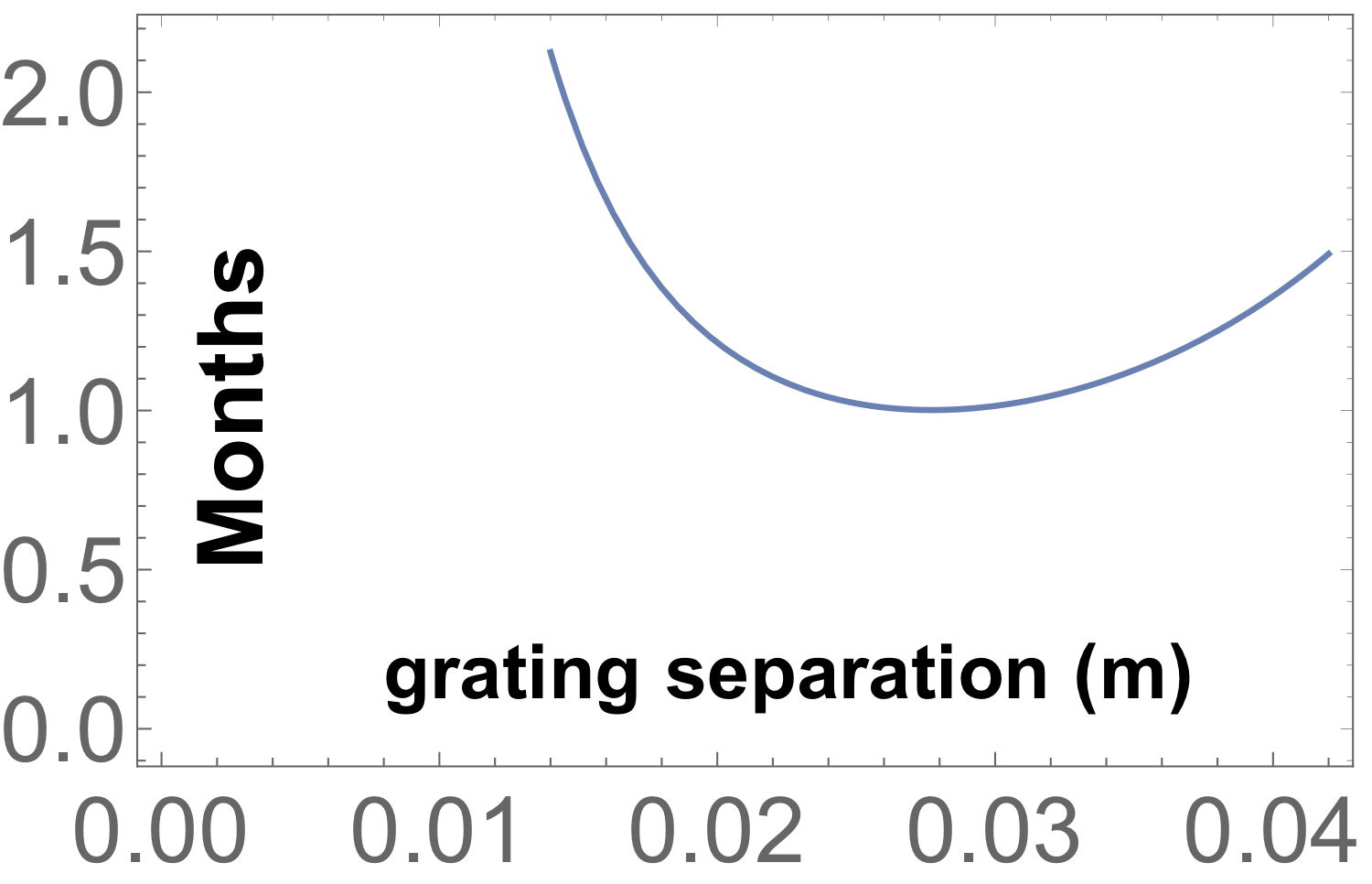}
\caption{``Thin-film'' SFHe beam}
\end{subfigure}~~~
\begin{subfigure}{0.33\textwidth}
~\includegraphics[width=.96\textwidth,trim=0 3 0 0,clip]{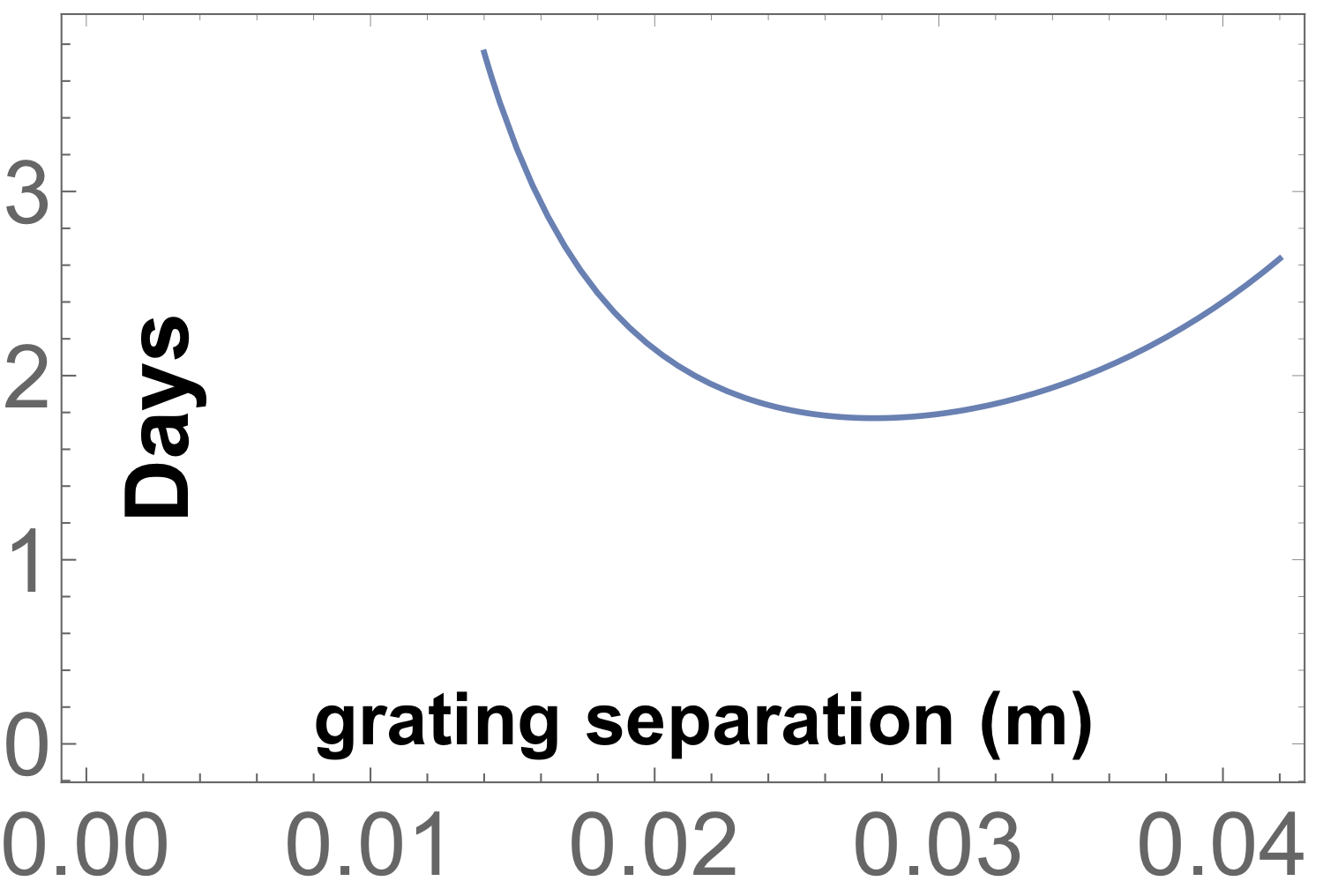}
\caption{``Thick-film'' SFHe beam}
\end{subfigure}
\vspace{-.2in}\caption{Representative MAGE sensitivity estimates vs.\ grating separation for beam options described in text, with 0.5\,$\mu$m-thick gratings of 100\,nm pitch, assuming 10\% contrast and that statistical uncertainties dominate over systematics; shown is beam time required for 5$\sigma$ determination of the sign of \gbar\ (i.e., $\delta\gbar/g=0.4$).$^8$}\label{fig:sens}
\end{figure}

\begin{figure}[tb]
\centerline{\includegraphics[width=.8\textwidth,trim=0 40 0 0 mm,clip]{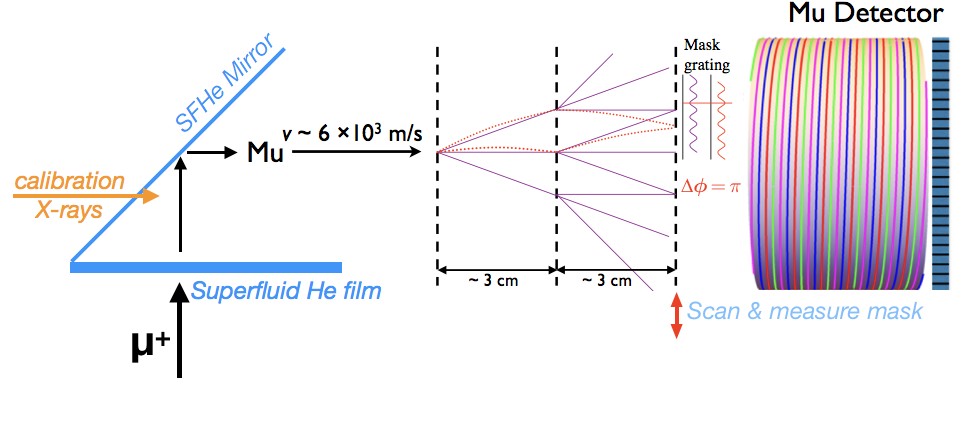}}
\caption{MAGE concept drawing}\label{fig:system}
\end{figure}

The \gbar\ sensitivity estimates given above (and in Fig.~\ref{fig:sens}) were calculated assuming the use of PSI's $\pi$E5 beamline, with $1.2\times 10^8$ surface muons/s, and 10\% interferometer contrast.\footnote{Note that the sensitivities of Fig.~\ref{fig:sens} are somewhat more pessimistic than that of Eq.~\ref{eq:sens}, due to inclusion of estimated decay losses from the Mu source to the first grating. The size of this effect will depend on the final source--interferometer distance, which will depend on cryostat engineering  details yet to be determined.} In both the thin- and thick-film approaches, the converter must be a flat, horizontal pool of SFHe at the bottom of a cryostat see Fig.~\ref{fig:system}). The resulting vertical Mu beam can be deflected into the horizontal (as required by the interferometer) by reflecting it off of a 45$^\circ$-inclined SFHe-coated surface~\cite{SFHe-Lens}. This has the further advantage that antimuons failing to stop in the SFHe layer
pass in front of, rather than through, the detectors, where they could otherwise be counted, potentially contributing to accidental background coincidences. It also frees up space for a calibration X-ray beam (discussed below).

\subsection{Interferometer Alignment and Calibration}
A key problem that must be solved is how to determine the zero-deflection interference phase.\footnote{The ``obvious'' solution of comparing muonium and antimuonium beams is unfortunately not feasible, since it is impractical to produce a sufficiently positron-rich $\mu^-$ stopping medium.} The solution we have adopted is to calibrate the interferometer using soft X-rays. This is most simply accomplished if the MCP used to detect the slow electrons can also serve as an X-ray detector;\footnote{If necessary, to eliminate the possible ambiguity between Mu and X-ray events, the $e^-$ accelerating potential can be turned off during calibration runs.} suitable MCPs are commercially available. An X-ray energy of 2\,keV gives the same wavelength as that of 6.3\,mm/$\mu$s muonium; this is an energy within the range of commercially available benchtop X-ray sources. At this wavelength Si$_3$N$_4$ is partially transparent, reducing undesired heating of the gratings by the calibration source; an interference pattern is nevertheless formed due to the combination of phase shift and partial absorption of the X-rays that pass through the grating material. If need be, the contrast can be enhanced by using a more absorbing grating in the third position. 
The undesirable heating is further reduced if the laser gauge is long-term stable such that it is unnecessary to repeat the X-ray calibration at short intervals.  Figure~\ref{fig:Allan} shows that the interval between X-ray calibrations can be about 15 minutes assuming we require the calibration to be good to 3\,pm. Reducing that requirement to 10 pm extends the recalibration interval to 4 hours based on an extrapolation of the TFG performance assuming it displays ``1/f noise.'' Figure~\ref{fig:optics} shows the laser interferometer operating in a cavity, which increases its performance. By similarly extrapolating the Allan deviation for a TFG operating in a cavity (finesse of 130)~\cite{Thapa-etal} and requiring the 3\,pm calibration, we get a recalibration interval of over ten days. This essentially eliminates the heating problem.

\section{Systematic Uncertainties}

Controlling and understanding systematic uncertainties will be critical to the success of the experiment.
In particular, it will be crucial to verify
that the measured deflection is in fact due to gravity. 
The interferometer
will therefore  be designed to be rotatable about the beam axis; with grating
lines vertical, there will be no gravitational phase shift. 
The interferometer will be constructed to rotate by  90$^{\circ}$ and 180$^{\circ}$, allowing  upright, inverted,  and sideways 
operation. Rotation will result in distortions of the support structure.
A 1-cm-tall silicon post supported at the bottom shortens under gravity
by 6--9\,pm, depending on crystal orientation. A cantilever deflects
by a larger distance. 
We address these distortions with several measures:
(1)~The optical bench (Fig.~\ref{fig:interf}a) is a short
and thick structure suspended at two points along the beam axis chosen to null the gravity effect on the critical grating-position combination $y_1+y_3-2y_2$ (where $y_i$ refers to the vertical position of grating $i$). (2)~We take advantage of a high degree of symmetry
in its construction and support. (3)~The grating bars themselves are
supported at $\sim$\,$\mu$m intervals with buttresses (fabricated
out of the same membrane), reducing their deflection  to a negligible level. (4)~The atom diffraction patterns are relatively narrow (first diffraction maximum at 5.6\,mrad), and we localize both Mu and X-ray events to $\sim$\,1 mm, which allows use of the X-ray interference pattern to monitor and correct for gravitationally induced changes due to inversion, as well as other possible effects that may depend on transverse position.

Electric- and magnetic-field gradients shift the observed Mu atom interferometric phase. The electric fields are caused by ``patch effect'' surface potential variations on the electrostatic shielding. Since the electrostatic shield inverts with the interferometer, the electric effect cancels well with interferometer inversion. For magnetic effects (which do not entirely cancel with inversion), the Kirch group at ETH/PSI can monitor and shield at the $<$\,nT/mm level~\cite{PhysRevLett.115.162502}, which is sufficient. Table~\ref{tab:syst} shows that with reasonable shielding measures these effects are a small fraction of 1\,g and below the level of concern.
The van der Waals effect (vdWE) can also cause an interferometric phase shift. A shift of $\approx$\,300\,mrad was observed by placing a Si$_3$N$_4$ grating in one arm of an atom interferometer~\cite{Perreault2005a} (cf.\ the 12\,mrad shift due to gravity in MAGE for a grating spacing of 2 lifetimes). Two factors make the vdWE larger in MAGE: the gratings may be thicker ($\sim$\,3-fold), and we anticipate using metallization on the upstream face ($\sim<$\,2-fold enhancement; see Table~I of \cite{Perreault2005b}). Several, more significant, factors reduce the vdWE in MAGE. (1)~Both interferometer beams see the same gratings, so vdWE- induced phase shift results only from inhomogeneous variation in grating structure. (2)~The Mu atom's static polarizability is similar to that of H, 34 times smaller than that of Na, implying that the effect on visibility in MAGE is much smaller than in \cite{Perreault2005a}. (3)~The vdWE phase shift cancels with interferometer inversion. Thus, the vdWE will be negligible in MAGE.

\begin{table}
\caption{Summary of gradient effects on \gbar.}\label{tab:syst}
\centering
\begin{tabular}{@{}lcccc@{}}\toprule
Cause & Size & Parameter & Value & Effect ($g$) \\
\midrule
$E$ gradient & 100\,V/m$^2$ & $\alpha_{\rm Mu}$ & $7\!\times\!10^{-31}$\,m$^3$ &  0.04\,$^*$\\
$B$ gradient & $<$1\,nT/mm &$\mu_{\rm Mu}$ & $9\!\times\!10^{-24}$\,J/T & $<$0.005\,$^\dagger$ \\
\bottomrule
\end{tabular}
\begin{itemize}\setlength\itemsep{-4pt}
\small
\item[ ]{\hspace{1.15in}$^*$~Cancels on inversion\qquad$^\dagger$~Partially cancels on inversion}
\end{itemize}
\end{table}
 
 A useful systematics test will be to measure, using various atom beams as well as muonium, the Sagnac effect due to the Earth's rotation, which depends on both the compass direction of the beam and the angle of the gratings relative to the horizontal (or vertical). For our geometry, beam velocity, and latitude, the effect is about 10\% as big as the gravitational deflection. This will demonstrate the capability of the interferometer to measure phase shifts smaller than that due to gravity. Operating the interferometer with atoms of normal matter also tests the fidelity of calibration using X-rays and inversion.

\section{Prospects}

An R\&D program on SFHe-produced Mu beams is proceeding at PSI in parallel with the muCool project. It is anticipated that such beams may be available at PSI within about 3 years. We hope to produce the interferometer and detectors in parallel, so as to be ready for the first muonium gravity measurements in the early 2020s.

\vspace{6pt} 

\acknowledgments{
We thank Alex Cronin of the University of Arizona,  Ben McMorran of the University of Oregon, and Grant Bunker, Yurii Shylnov, and Jeff Terry of IIT for useful discussions, as well as IIT undergraduate Brinden Carlson for help with TFG tests. 
This work supported by the Swiss National Science Foundation (under grants 200020\_172639 and 206021\_170734) and the ETH Z\"urich (under grant ETH-35 14-1) as well as the IIT IPRO program~\cite{IPRO} and Physics Dept., College of Science.}

\authorcontributions{All authors contributed equally to this work.
}

\conflictsofinterest{
The authors declare no conflict of interest.} 


\abbreviations{The following abbreviations are used in this manuscript:\\

\noindent 
\begin{tabular}{@{}ll}
CNM & the Center for Nanoscale Materials at Argonne National Laboratory\\
FWHM & full-width at half maximum\\
\gbar, $g$ & the gravitational acceleration at the earth's surface of antimatter and matter, respectively\\
GR & general relativity\\
IIT & Illinois Institute of Technology\\
IPRO & Inter-Professional Project\\
IR & infrared\\
MAGE & the Muonium Antimatter Gravity Experiment\\
MCP & microchannel plate\\
MDPI & Multidisciplinary Digital Publishing Institute\\
Mu & muonium \\
muCool & cooled muon beam R\&D program at PSI\\
PBS & polarizing beam splitter\\
PSI & Paul Scherrer Institute\\
RMS & root-mean-square\\
SFHe & superfluid helium\\
TFG & tracking frequency gauge\\
UNCD & ultrananocrystalline diamond\\
vdWE & van der Waals effect\\
\end{tabular}}


\end{document}